\headline{\hfill }
\font\lr=cmr10 scaled 1440
\magnification=\magstep1
\baselineskip 14pt

\def\au{{\rm AU}}
\def\etal{{\it et al.}}

\def\bV{{\bf V}}
\def\kms{{\rm km/s}}

\def\gtsim{\mathrel{\raise.3ex\hbox{$>$}\mkern-14mu
             \lower0.6ex\hbox{$\sim$}}}
\def\ltsim{\mathrel{\raise.3ex\hbox{$<$}\mkern-14mu
             \lower0.6ex\hbox{$\sim$}}}
\def\gtorder{\gtsim}
\def\ltorder{\ltsim}


\null\bigskip
\centerline{\bf Binary-single-star scattering --- VII.}
\centerline{\bf Hard Binary Exchange Cross Sections for Arbitrary Mass Ratios:}
\centerline{\bf Numerical Results and Semi-Analytic Fits}
\bigskip
\centerline{D.~C.~Heggie}
\smallskip
\centerline{University of Edinburgh,}
\centerline{Edinburgh EH9 3JZ, U.K.}
\centerline{\it d.c.heggie@ed.ac.uk}
\medskip
\centerline{P.~Hut}
\smallskip
\centerline{Institute for Advanced Study,}
\centerline{Princeton, NJ 08540}
\centerline{\it piet@sns.ias.edu}
\medskip
\centerline{S.~L.~W.~McMillan}
\smallskip
\centerline{Department of Physics and Atmospheric Science,}
\centerline{Drexel University,}
\centerline{Philadelphia, PA 19104}
\centerline{\it steve@zonker.drexel.edu}
\bigskip
\bigskip\noindent

\centerline{\bf Abstract}
\medskip\noindent
We present the first comprehensive fitting formula for exchange
reactions of arbitrary mass ratios.  In a comparison with numerical
results, this expression is shown to be accurate in the hard binary
limit to within 25\% for most mass ratios.  The result will be useful
in forming quantitative estimates for the branching ratios of various
exchange reactions in astrophysical applications.  For example, it can
be used to construct quantitative formation scenarios for unusual
objects in globular clusters, such as binaries containing a pulsar.
\smallskip

\par\vfill\eject\noindent
{\bf 1. Introduction}
\medskip 

In stellar dynamics, an exchange reaction is a particular type of
interaction between a binary star and an incoming third star, where
one of the components of the binary is expelled and its place is taken
by the incomer.

In the long history of the three-body problem, the study of exchange
reactions had a curious start.  Though numerical examples were given
by Becker (1920) long ago, it appears that the very {\it possibility}
of such reactions remained controversial until later numerical work
published in 1975 (Marchal 1990).  This is all the more curious
because in that very year Hills (1975) and Heggie (1975) separately
published different treatments which, taking the existence of exchange
for granted, attempted to determine how its probability depended on
such parameters as the initial speed of the incomer, and the masses of
the stars.

Since that time a considerable number of studies of exchange reactions
have been carried out.  In this paper we concentrate on the problem of
determining the {\it cross section} for exchange, in the limit in
which the initial speed of the incoming star is very low, but for all
possible masses.  Our study will therefore be largely complementary to
previous work, which has often adopted different restrictions,
e.g. encounters at zero impact parameter (Hills \& Fullerton 1980),
the case in which one component of the binary has very low mass (Hills
\& Dissly 1989), or the initial eccentricity is zero (Hills 1991, 1992).  
Other work of this kind will be mentioned in \S5.1, for
comparison with our own data.

This paper is the seventh in a series discussing many aspects of
three-body scattering in the point mass approximation (Hut \& Bahcall
1983, Hut 1983, 1993, Heggie \& Hut 1993, Goodman \& Hut 1993,
McMillan \& Hut 1996), but this is the first to deal with stars of
unequal mass.  For the case of equal masses, much further information
on exchange cross sections will be found in earlier papers of this
series, especially Papers II and IV, as well as the atlas of hard
binary scattering cross sections provided by Hut (1984).

The present paper is arranged as follows.  In the following section we
describe the numerical software we have used for generating cross
sections.  Because of its highly automated yet flexible construction,
this is a topic of interest in its own right.  In \S3 we analyse the
problem analytically, in order to understand the dependence of the
exchange cross section on the masses, especially in various asymptotic
regimes.  It turns out to be possible to write down a single
expression which accommodates all asymptotic regimes.  Section 4
synthesises all our numerical data and asymptotic theory to provide a
comprehensive and simple formula which is believed to be approximately
valid for all masses.  In the concluding section it is tested against
previous work by Sigurdsson \& Phinney (1993), Hills (1992), and
Rappaport \etal\ (1989), and then we summarise our results.

\bigskip
\noindent
{\bf 2. Numerical Computation of Exchange Cross Sections}
\medskip\noindent
{2.1 Software for Three-Body Scattering}
\smallskip\noindent

The first sets of binary---single-star scattering experiments were
reported by Hills (1975) and Heggie (1975).  In the former, most
encounters took place at zero impact parameter.  The first direct
determination of accurate cross sections and reaction rates for
binary---single-star scattering was made by Hut \& Bahcall (1983).
For each type of total or differential cross section, a detailed
search of impact parameter space was performed as a pilot study,
before production runs were started.  The problem with the choice of
impact parameter (lateral offset from a head-on collision, as measured
at infinity) is this: allowing too large an impact parameter can imply
a large waste of computer time on uninteresting orbits; while choosing
too small an impact parameter will yield a systematic underestimate of
some cross sections, since some encounters of interest will be missed.

The first {\it automatic} determinations of cross sections and
reaction rates for binary---single-star scattering are described by
McMillan \& Hut (1996; hereafter referred to as Paper VI).  Rather
than relying on human inspection of pilot calculations, their software
package includes an automatic feedback system that ensures
near-optimal coverage of parameter space while guaranteeing
completeness.  We refer the interested reader to Paper VI for further
details on the STARLAB software package.  References to earlier papers
on 3-body scattering can be found in the recent papers by Hills
(1992), Heggie \& Hut (1993), Hut (1993) and Sigurdsson \& Phinney
(1993).

\medskip\noindent
{2.2 Numerical Results}
\smallskip
First we explain our notation.  Let $m_1$ and $m_2$ be the masses of
the components of the binary, and $m_3$ that of the incoming third
star.  Then we define $M_{12} = m_1+m_2$, $M_{123} = m_1+m_2+m_3$,
etc.  When the incoming third body and the binary are still at a very
large distance, let the semi-major axis of the binary be $a$ and the
relative speed of the third body and the barycentre of the binary be
$V$.  Then we can scale $V$ by the critical value, $V_c$, for which
the total energy of the triple system, in the rest-frame of its
barycentre, is zero, i.e. let
$$
v = V/V_c,\eqno(1)
$$ 
where  
$$
V_c^2 =
Gm_1m_2M_{123}/(M_{12}m_3a).\eqno(2)
$$
All of our runs have been carried out with $v = 0.1$.

We also scale the cross sections themselves, in two ways.  Let
$\Sigma$ be the cross section for an exchange process.  First,
following Paper I in this series, we may scale out the gravitational
focusing of the third body as it approaches the binary, and also the
physical cross section of the binary itself, by defining the
dimensionless cross section
$$
\sigma = {v^2\Sigma\over\pi a^2}.\eqno(3)
$$  
This definition has the disadvantage that $\sigma$ becomes very large
when $m_3\gg M_{12}$, and we have found it convenient to scale $V$ not
by $V_c$ but by a typical speed reached by the third body when it
makes a close approach to the binary.  To be precise, we define a
speed $V_g$ by $\displaystyle{V_g^2 = {GM_{123}\over2a}}$, and let
$\bar v = V/V_g$.  Evidently $V_g$ is the relative speed of the third
body if it falls from rest at infinity to a distance $4a$ from a body
of mass $M_{12}$.  We have chosen this slightly odd numerical factor
so that $V_g = V_c$ in the case of equal masses.  Thus we define a new
dimensionless cross section $\bar\sigma = V^2\Sigma/(\pi a^2V_g^2)$,
i.e.
$$
\bar\sigma = {2V^2\Sigma\over\pi GM_{123}a},\eqno(4)
$$
which is related to the previous definition by  $\bar\sigma =
\sigma\bar v^2/v^2$.

The results of our runs are displayed in Table 1, with estimates of
$1$-$\sigma$ errors.  In general we have attempted to ensure that the
total exchange cross section for each component of the binary is
calculated to better than about ten percent, though there are clearly
cases in which the cross section must be so small that the resulting
computational effort would be prohibitive.  This is also illustrated
in Fig.1, whose main purpose is to show the coverage of the plane of
mass ratios in our numerical experiments.  We covered all mass ratios
up to a maximum of 2dex (between any pair of stars in the system), in
steps of 0.5dex, as well as a few other cases.

It is helpful to distinguish two different kinds of exchange reaction
(e.g. Heggie 1975), and Table 1 presents results for both.  In {\it
direct} exchange the encounter terminates promptly, and the orbits are
uncomplicated; while in {\it resonant} exchange the three bodies form
a temporary bound system, and the escaping particle emerges only after
several interactions.  The distinction between direct and resonant
encounters is not always a clear one; the operational procedure used
to classify our numerical results is presented in Paper VI.

A few other remarks about the data in Table 1 should be made at this
point. First, for a few mass ratios we have data from additional runs
which are not shown here.  Those shown are the data sets with the
smallest errors.  The additional data has been used, however, in the
parameter fitting in \S4 below.  Second, entries with a star indicate
experiments in which no events of the relevant kind were observed.
Though it might be thought that it should be possible, on the basis of
a large number of scattering experiments, to give an {\it upper bound}
for the cross section of a process which produced no events, this is
actually not rigorously possible because of the organisation of the
software, which decides on the range of impact parameters on the basis
of {\it observed} events.  If the software has no evidence on the
range of impact parameters which can produce a given event then it is
possible that it can miss a range where the process is important.
Therefore this null data should simply be discarded.

\bigskip\noindent
{\bf 3. Asymptotic Theory of Exchange Cross Sections}
\bigskip

In this section we address theoretically the problem of determining
cross sections for the reactions discussed in this paper, i.e.
exchange for hard binaries.  Our aim will be to determine the way in
which they scale with the parameters of the problem, especially in the
extreme regimes of masses.  Some of these extreme cases might seem
physically implausible or unimportant, but the purpose of the theory
is to try to account for trends in the numerical data, and to suggest
ways in which the numerical results might be extrapolated.

In general we consider the approach of a third body whose speed ``at
infinity'' relative to the barycentre of the binary was $V$, and denote
the initial semi-major axis of the binary by $a$.  We shall label the
components of the binary such that $m_1$ is the mass of the component
which is ejected as a result of the encounter, and in this section we
shall generally add a subscript 1 to symbols for the cross section, in
order to reinforce this convention.  It must always be borne in mind
that we are dealing with the case in which $V$ is very small.

\medskip\noindent
3.1 The Case of a Massive Incoming Star.
\medskip

First we consider the regime in which both $m_1\ll m_3$ and $m_2\ll
m_3$.  In this case the incomer is very massive, and we shall see that
it is possible for a tidal encounter by the third body to unbind the
binary.  Since we are always considering the hard binary limit
(i.e. $V \ll V_c$), the three stars cannot escape singly to infinity,
and so an exchange interaction must occur.  What is less obvious is to
decide which component escapes.

Let $R_p\gg a$ denote the distance of closest approach of the third
body.  (The subscript denotes pericentre.)  At this distance its speed
relative to the barycentre of the binary is denoted by $V_p$ and can
be estimated from
$$
V_p^2\sim GM_{123}/R_p.\eqno(5)
$$
The duration of the encounter is of order $R_p/V_p$.  During this
period the (tidal) acceleration of the relative motion of the binary
components by the third body is of order $Gm_3a/R_p^3$.  Therefore the
change in their relative speed is of order $\Delta V_{12}\sim
Gm_3a/(R_p^2V_p)$.  This is enough to disrupt the binary if $(\Delta
V_{12})^2\gtorder GM_{12}/a$, i.e. if
$$
R_p^3\ltorder m_3^2a^3/(M_{123}M_{12}).\eqno(6)
$$
Incidentally we have assumed implicitly in all this that the duration
of the encounter does not much exceed the period of the binary, for
then the change in energy of the binary would be exponentially smaller
than the estimate we have used (Heggie 1975).  In fact it is easy to
show that the assumption on time scales is justified {\it post hoc} by
eq.(6).

If, as we assume throughout, the original speed of approach ($V$) of
the third body is very small, the barycentre of the binary moves along
a nearly parabolic orbit relative to $m_3$.  As a result of the
disruption of the binary, one component will now be moving in front
of, and faster than, the barycentre, while the other will fall behind,
and the probability that a given component is the one that moves in
front is roughly the same for both components.  Call this body $m_f$ and
the other component $m_b$.
If $m_f\gg m_b$, then $m_f$ is moving only slightly faster than the
barycentre, but since the barycentre is moving on a slightly
hyperbolic orbit, it follows that $m_f$ will escape.  If, on the other
hand, $m_f\ll m_b$, then $m_f$ will pick up almost all the extra
velocity ($\Delta V_{12}$) generated in the encounter, which is enough
to ensure its escape.  In either case, then, $m_f$ is the component of
the original binary which escapes, and, by the convention already
stated, we identify this with $m_1$.

It might be surprising to find that the probability that a given
component is the escaper does not diminish significantly when it is
much more massive than the other component.  This is, however,
confirmed in Table 1 by such cases as $m_1 = 0.09$, $m_3 = 9.09$.

It remains only to estimate the cross section for this process. The
impact parameter of the third body, $p$, when at a large distance on
its incoming orbit, is related to other parameters of the encounter by
approximate angular momentum conservation in its motion relative to
the barycentre of the binary, which leads to $pV = R_pV_p$.  Therefore
the cross section for these encounters is $\Sigma_1\sim p^2\sim
(R_pV_p/V)^2$.  (The factor $(V_p/V)^2$ accounts for the
``gravitational focusing'' that takes place as the third body
approaches on its nearly parabolic orbit.)  Using eq.(5), and the
expression on the right of eq.(6) to estimate $R_p^3$, we find that
$\Sigma_1 \sim
\displaystyle{{GM_{123}a\over V^2}\left({m_3^2\over
M_{123}M_{12}}\right)^{1/3}}$.

In view of our assumptions that $m_1$ and $m_2$ are very small
compared with $m_3$, we see that an equally valid asymptotic
expression is
$$
\Sigma_1 \sim
\displaystyle{{GM_{123}a\over V^2}\left({M_{123}\over
M_{12}}\right)^{1/3}}.\eqno(7)
$$ 
This will be more useful for combining with other asymptotic formulae
later.

Finally we have to dispose of the question of resonant exchange.  In
fact, if, after this first encounter, the third body is bound to the
binary without disrupting it, thus forming a resonance, then on each
subsequent encounter it is quite likely to become unbound again, and
so the cross section for resonant exchange cannot greatly exceed that
just estimated for direct exchange.  Again this is confirmed by data
in Table 1, though it is noticeable that the cross section for
ejection is now somewhat larger for the less massive star.

This discussion has been taken fairly slowly so as to introduce the
various steps and considerations involved in the estimate of the
exchange cross section in a certain regime.  In subsequent discussions
only the distinctively different aspects will be treated in comparable
detail.

There are several aspects of this kind of encounter which will be
useful later (\S3.2). The energy given to the binary will be
comparable to the energy required to break it up.  Therefore the
energy extracted from the relative motion of $m_3$ and the barycentre
of the binary must also be comparable with $Gm_1m_2/a$, and so this
can be taken as an estimate of the binding energy of the new binary.
It follows that the semi-major axis of the new orbit, denoted by
$a_{23}$, may be estimated by
$$
Gm_1m_2/a\sim
Gm_3m_2/a_{23},\eqno(8)
$$ 
i.e. $a_{23}\sim a m_3/m_1$.  Also the pericentric distance of the new
binary is of order $R_p$, and so can be expressed as
$$
R_p \sim a_{23}(m_1/m_3)(M_{123}/M_{12})^{1/3}.\eqno(9)
$$
It follows that the eccentricity of the new binary ($e^\prime$) is
high, since $1-e^\prime\sim R_p/a_{23}$.  Finally, the initial
separation of $m_1$ and $m_2$ is related to the new semi-major axis by
$$
R_{12}/a_{23}\sim (m_1/m_3).\eqno(10)
$$

The argument that the binding energy of the new binary must be
comparable with that of the original one has analogues in other
capture processes.  One familiar example is tidal capture in globular
star clusters (Fabian \etal\ 1975), where the encounter must be close
enough to remove the kinetic energy, $T$, of relative motion of the
two stars.  In this case it follows that the binding energy of the new
binary will be comparable to $T$.  The analogy is not precise,
however, because in this case the binding is not achieved by
disruption of one of the stars.

\medskip\noindent
3.2 The Case of Ejection of a Massive Star.
\smallskip
We move on now to the regime in which $m_2\ll m_1$ and $m_3\ll m_1$
The first point to notice about this case is that it may be obtained
from the previous case by time-reversal.  In the previous case a heavy
incoming third body becomes a component in the new binary, while in
the present case it is the heavy component which is ejected.  (Recall
that the components of the binary are named in such a way that it is
the first component which is exchanged.)  While the ejection of the
most massive component in a three-body interaction seems unlikely, the
use of time-reversal yields the special circumstances in which it can
happen.  In fact the encounter must occur while the two components of
the binary are close to pericentre (since $R_p \ll a_{23}$ in the
previous case), and the incoming star must come even closer to the
lighter component, $m_2$.

We shall return to some aspects of a direct calculation of the cross
section towards the end of this section.  Mainly, however, we shall
exploit the fact that time-reversal leads to a process whose cross
section we have already estimated, and so we may use the theory of
detailed balance to estimate the required result.  This theory is set
out in the Appendix, and leads to the following result for inverse
process.  (Here we omit  the subscript 1 on the cross section, since the
identity of the escaper is sufficiently defined by other aspects of the
notation.)

Let $\displaystyle{{d\Sigma\over dE_{23}}(E_{12}\to E_{23})}$ be the
differential cross section for the formation of a binary having
components $m_2$, $m_3$ and energy $E_{23}$, from a binary having
components $m_1$, $m_2$ and energy $E_{12}$, by exchange.  Then this
is related to the differential cross section for the inverse process
by
$$
{d\Sigma\over dE_{12}}(E_{23}\to
E_{12}) = \left({m_1\over m_3}\right)^{5/2}\left({M_{12}\over
M_{23}}\right)^{1/2}{V_3^2\over V_1^2}\left({E_{12}\over E_{23}}\right)^{-5/2}
{d\Sigma\over dE_{23}}(E_{12}\to
E_{23}).\eqno(11)
$$
Here $V_1$ and $V_3$ are the speeds of the incoming single body in the
two scatterings.  The coefficient on the right side comes from the
phase space volumes associated with the reactants: for the incoming
single body this is proportional to $V_i^2$, and the factor
$E_{ij}^{-5/2}$ is easily understood in a similar way in terms of the
internal degrees of freedom of the binary (Hut 1985).

This relation involves differential cross sections, whereas eq.(7)
estimates a total cross section.  However, we already estimated that
the typical energy of the new binary in that case is $E_{23}\sim
E_{12}$, and if we estimate $\displaystyle{{d\Sigma\over
dE_{12}}}(E_{23}\to E_{12}) \sim \Sigma(E_{23}\to E_{12})/E_{12}$, we
see that eq.(11) yields the following estimate for the total cross
section:
$$
\Sigma(E_{23}\to
E_{12}) \sim \left({m_1\over m_3}\right)^{5/2}\left({M_{12}\over
M_{23}}\right)^{1/2}{V_3^2\over V_1^2}\Sigma(E_{12}\to
E_{23}).
$$
Substituting eq.(7) in the right side we find, for the case $m_1$,
$m_2\ll m_3$, the estimate
$$
\Sigma(E_{23}\to
E_{12}) \sim {GM_{123}a_{12}\over V_1^2}\left({M_{123}\over M_{12}}\right)^{1/3}\left({m_1\over m_3}\right)^{5/2}\left({M_{12}\over
M_{23}}\right)^{1/2},
$$
where $a_{12}$ (heretofore denoted simply by $a$) is the semi-major
axis of the binary with components 1 and 2.  Now we use eq.(8) to
replace $a_{12}$ by $a_{23}m_1/m_3$, and interchange the labeling of
stars $1$ and $2$ to restore our customary labeling of the
incomer. This yields the exchange cross section
$$
\Sigma(E_{12}\to
E_{23}) \sim {GM_{123}a_{12}\over V_3^2}\left({M_{123}\over M_{23}}\right)^{1/3}\left({m_3\over m_1}\right)^{7/2}\left({M_{23}\over
M_{12}}\right)^{1/2}
$$
in the case $m_2$, $m_3\ll m_1$.  An equally valid asymptotic formula
is obtained by replacing $M_{12}$ by $M_{123}$, and then we see that a
formula which is compatible with eq.(7), and therefore valid in both
the asymptotic regimes of \S\S3.1 and 3.2, is
$$
\Sigma_1\sim {GM_{123}a\over V^2}\left({M_{23}\over M_{123}}\right)^{1/6}\left({m_3\over M_{13}}\right)^{7/2}\left({M_{123}\over
M_{12}}\right)^{1/3}.\eqno(12)
$$

Before we leave this regime it is worth noting that the major part of
the mass-dependence here, i.e. the factor $(m_3/ M_{13})^{7/2}$ is
easily understood. We already saw (eq.(9), but with relabeling
appropriate to time-reversal) that the separation of the binary
components at the time of approach of $m_3$ must be of order
$R_{12}\sim a (m_3/m_1)(M_{123}/M_{23})^{1/3}$.  Now the probability
of this (for a thermal distribution of binaries of a given energy) is
of order $(R_{12}/a)^{5/2}$ when $R_{12}\ll a$ (cf. Paper IV).  Next,
the distance of closest approach of $m_3$ to $m_2$ must be of order
$R_p\sim a m_3/m_1$, by eq.(10), and at this time the speed of the
third body, which is gained mostly by falling to within a distance
$R_{12}$ of $m_1$, is given by $V_p^2\sim GM_{123}/R_{12}$.  It
follows that the cross section is
$\Sigma_1\sim\displaystyle{{R_p^2V_p^2\over V^2}\left({R_{12}\over
a}\right)^{5/2}},$ $\sim\displaystyle{{GM_{123}a\over
V^2}\left({m_3\over m_1}\right)^{7/2}\left({M_{123}\over
M_{23}}\right)^{5/6}}$.  This is larger than our estimate in eq.(12)
because we have not taken into account the special circumstances of
the encounter which allows $m_3$ to be captured by $m_2$, but does
explain the major part of the mass dependence: all three stars must
come within a separation which is of order the pericentric distance of
a binary of high eccentricity.  The factor $(m_3/m_1)^{7/2}$ is a
measure of the phase space volume available to such systems.

\medskip\noindent
3.3 The Case in Which a Massive Component Remains
\smallskip

Finally we turn to the regime in which both $m_1\ll m_2$ and $m_3\ll
m_2$.  We first discuss the case in which $m_3\gg m_1$, i.e. an object
of intermediate mass displaces a low-mass companion of an object of very
high mass. (Recall our convention for labelling the components, which is
that $m_1$ is the mass of the component which is ejected.) To begin, let
us consider the possibility of direct exchange.  Suppose $m_3$
approaches $m_1$ within a distance $R_p\ll a$.  Then the speed of $m_3$
will be given by $V_p^2\sim GM_{123}/a$ provided that the influence of
$m_1$ is not dominant, i.e. provided that $R_p\gtorder aM_{13}/M_{123}$.
It follows that the speed imparted to $m_1$ is of order the escape speed
provided that $R_p\sim a M_{13}/M_{123}$.  Hence the cross section for
direct exchange is $$ \Sigma\sim {GM_{123}a\over V^2}\left({m_3\over
m_2}\right)^2.\eqno(13) $$

Now suppose only that the distance of closest approach between $m_3$
and $m_1$ is of order $a$.  Then there is a significant probability
that $m_3$ will become bound to the binary without ejecting $m_1$,
i.e. a resonance will form.  Thus the cross section for resonance {\it
formation} greatly exceeds the cross section for direct exchange,
since our estimate of $R_p$ for direct exchange requires that $R_p\ll
a$.  In fact the cross section for formation of a resonance is simply
$$
\Sigma_1\sim  {GM_{123}a\over V^2}.\eqno(14)
$$

At this point in the evolution of the resonance the binding energy of
$m_3$ is of order the change in energy of $m_1$, which is of order
$Gm_1m_3/a$.  Note that this estimate is valid no matter how small
$m_1$ is; we are always considering the limit of extreme hardness, and
even a small change in the energy of the third body can bind it to the
binary if its energy at infinity was sufficiently small.  On the other
hand, the cross section for formation of a resonance leading to
exchange cannot greatly exceed our estimate: if the closest distance
of approach of $m_3$ greatly exceeds $a$ then a hierarchical triple
system forms, and exchange is very improbable.

Now we must estimate the probability that the resonance will be
resolved with the escape of star $m_1$. We think of the binding energy
of this particle, $E_1$, as performing a random walk under the
influence of repeated passages by star $m_3$.  The typical change in
$E_1$ is of order $Gm_3m_1/a$.  Because we are assuming that $m_3\gg
m_1$, it might be thought that the mean effect would be to
systematically unbind $m_1$, by a kind of mass segregation.  However,
the mean change in $E_1$, taken over an ensemble of such systems, is
actually of second order in the ratio of the masses: in the
approximation of first-order perturbation theory, time-reversal shows
that for each change in $E_1$ there is a system in which the change
has the opposite sign.  Therefore we shall assume that the mean change
for a given system is negligible.

As an aside, it is worth mentioning here that the system is a kind of
hierarchical triple.  Usually this term is used in reference to a
binary about which a third star revolves on a large elliptical orbit
which is well separated spatially from the binary.  In that case the
perturbation of the third body is weak because the orbit of the third
body is large.  In the present case there is no such spatial
separation, but still the perturbation of the third body by the binary
is weak, and this is due to the low mass of one component of the
binary.

We must estimate the probability that $E_1$, starting from the value
of order $Gm_1m_2/a$, may randomly walk, in steps of order
$Gm_1m_3/a$, to the value 0 without first reaching the value
$Gm_1m_2/a + Gm_1m_3/a$ (as $m_3$ would then escape again).  Now the
probability that a one-dimensional random walk exits from a given
boundary is a linear function of the initial position, varying from
unity at the boundary of interest to 0 at the opposite boundary.  In
the present case the boundaries are at $E_1 = Gm_1m_2/a + Gm_1m_3/a$
and $0$, and so the probability of escape at the boundary $E_1 = 0$,
starting at $E_1 = Gm_1m_2/a$, is of order $m_3/m_2$.  This can be
estimated equally well as $M_{13}/M_{123}$, and so it follows, using
eq.(14), that the cross section for resonant exchange is
$$
\Sigma_1\sim {GM_{123}a\over V^2}\left({M_{13}\over M_{123}}\right).
$$
It also follows that we obtain a form which is asymptotically correct
in all regimes studied so far if we modify eq.(12) to
$$
\Sigma_1\sim {GM_{123}a\over V^2}\left({M_{23}\over M_{123}}\right)^{1/6}\left({m_3\over M_{13}}\right)^{7/2}\left({M_{123}\over
M_{12}}\right)^{1/3}\left({M_{13}\over M_{123}}\right),\eqno(15)
$$
or, if everything is normalised by the total mass of the triple
system,
$$
\Sigma_1\sim {GM_{123}a\over V^2}\left({M_{23}\over M_{123}}\right)^{1/6}\left({m_3\over M_{123}}\right)^{7/2}\left({M_{123}\over
M_{12}}\right)^{1/3}\left({M_{13}\over M_{123}}\right)^{-5/2}.
$$
We also note that the energy of the new binary will be comparable with
that of the initial binary.

Observe that our estimate for resonant exchange is indeed larger than
the estimate, eq.(13), for direct exchange.  The fact that this is
true in the mass regime under discussion is also illustrated in Table
1 by such cases as $m_1=0.01,$ $m_3=0.099$.

Finally we turn to the case $m_3\ll m_1\ll m_2$, which is the time
reversal of the case just considered.  Though the foregoing argument
for the formation of a resonance goes through, it is now very unlikely
to be resolved by exchange.  We can, however, estimate the cross
section for resonant exchange by detailed balance, using the same
method as in \S3.2.  The result is that
$$
\Sigma\sim {GM_{123}a\over V^2}\left({m_3\over
m_1}\right)^{7/2}\left({M_{23}\over M_{12}}\right)^{1/2}\left({M_{13}\over M_{123}}\right),
$$
and we see that eq.(15) is, once again, a valid asymptotic result in
this regime.  {\it We therefore adopt eq.(15) as a cross section whose
form is valid in all regimes.}
Henceforth we drop the subscript 1 on $\Sigma$, and remind the reader
that $m_1$ is the mass of the component of the original binary which is
ejected.

\bigskip\noindent
{\bf 4. Synthesis of Numerical and Analytical Results}
\medskip\noindent
4.1 A Test of the Asymptotic Formulae
\smallskip
Though we already made a number of qualitative remarks relating the
theory of \S3 to the numerical data in Table 1, it is now time for a
more quantitative study.  Our aim is not yet to provide a fitting
formula for the data (which we take up in \S4.2), but nevertheless we
shall generalise eq.(15) slightly.  Using the notation of eq.(4), we
may rewrite eq.(15) as
$$
\bar\sigma\sim \left(1 + {m_1\over M_{23}}\right)^{-1/6}\left(1 +
{m_1\over m_3}\right)^{-7/2}\left(1+{m_3\over
M_{12}}\right)^{1/3}\left(1 + {m_2\over M_{13}}\right)^{-1}.
$$
In order to allow different multiplicative constants in the different
asymptotic regimes, we generalise this to
$$
\bar\sigma\sim a_1\left(a_2+{m_3\over
M_{12}}\right)^{1/3}\left(a_3 +
{m_1\over m_3}\right)^{-7/2}\left(a_4 + {m_1\over
M_{23}}\right)^{-1/6}\left(a_5 + {m_2\over M_{13}}\right)^{-1},\eqno(16)
$$
where the $a_i$ are constants.  There is then one asymptotic form for
the regime of \S3.1, one for that of \S3.2, and one each for the two
regimes in \S3.3 (i.e. the regimes in which $m_1\ll m_3$ and $m_1\gg
m_3$).

The formula in eq.(16) has been fitted to the {\it logarithm} of the
data in Table 1 using least squares, the standard deviation in the
logarithm being estimated by the relative error.  In cases where no
cross section was measurable the data point was assigned zero weight.

Since the aim of this exercise was to detect possible errors in the
theoretical asymptotic form we searched for deviations (between the
results of this formula and the experimental data) which were
significant (``$2$-$\sigma$''), large (above 1 in the natural logarithm),
and in an extreme mass regime.  We illustrate the results by
considering in a little detail the set of data in which the
discrepancies were largest.  This was the case $m_2\ll m_3\simeq
m_1/3$, as illustrated by Table 2.  The numerical results show a
trend, but not a significant one, and a constant value in the
numerical data (as predicted by the formula) is not ruled out at the
20\% level.  A very similar set of results, but with smaller
discrepancies and in the opposite sense, is found in the case $m_2\ll
m_1=m_3$.  Apart from these, the only systematic, large discrepancies
occur at one or two data points where the masses are comparable, and
the asymptotic formulae need not apply.  In conclusion, then, there is
no evidence that the theoretical formula is inconsistent, in the
appropriate regimes, with trends in the numerical data.

\bigskip\noindent
4.2 A Semi-Numerical Fitting Formula
\medskip
The theoretical results of \S3 are intended to provide the asymptotic
dependence of the exchange cross section on the masses of the
participants, but do not even attempt to provide numerical
coefficients for these.  The numerical data, on the other hand, apply
to only discrete points in the parameter space of the masses.  An
obvious way of synthesising the two kinds of result is to adopt a form
with the same asymptotic properties as the analytical result, but with
additional terms which are chosen to optimise the agreement with the
numerical data.  In a sense this is what was done in the previous
section, but with a limited degree of flexibility.  Here we adopt a
more general approach which could be extended more or less
arbitrarily.

Again we switch to
$\bar\sigma$, defined by eq.(4), and generalise  eq.(15) to
$$
\bar\sigma =  \left({M_{23}\over M_{123}}\right)^{1/6}\left({m_3\over M_{13}}\right)^{7/2}\left({M_{123}\over
M_{12}}\right)^{1/3}\left({M_{13}\over M_{123}}\right)\exp\left(\sum_{m,n;
m+n=0}^{m+n=N} a_{mn}\mu_1^m\mu_2^n\right),\eqno(17)
$$
where the $a_{mn}$ are constants, 
$$
\mu_1 = m_1/M_{12},~{\rm and}~ \mu_2 = m_3/M_{123}.\eqno(18)
$$
These two parameters entirely span the possible ranges of mass ratios
in the unit square $0\le\mu_1,~\mu_2\le1$.  The exponential is used to
constrain the function to be positive, and also to avoid altering the
asymptotic character of the leading expression.  By taking larger
values of the highest power $N$, it would be possible, in principle,
to improve the fit to arbitrary accuracy.

We have fitted formulae with $N \le 5$, and at the largest value the
value of $\chi^2$ is 133, which, considering the number of degrees of
freedom, is still rather large. (The number of data points is 126, and
the number of free coefficients is $21$.)  Nevertheless we suggest the
use of a cubic polynomial in the exponential in eq.(17), with the
coefficients given in Table 3.  As these 10 coefficients are quoted to
2 decimal places, the maximum relative error in the evaluation of
$\bar\sigma$ is about 5\%.  This is adequate in view of the accuracy
of the fit, which is discussed further below.

One worry about using a single polynomial for any kind of
interpolation is the possibility of large oscillations between data
points, but in fact the polynomial with the above coefficients is well
behaved.  Its range is about 5, i.e $\bar\sigma$ varies by a factor of
about 100 by the effect of this polynomial.  Its most noticeable
feature is a minimum value at $\mu_1=1$, $\mu_2=0$, i.e. the regime
$m_1\gg M_{23}$.  The cross section is very small in this region
anyway, because of the factor $(m_3/M_{13})^{7/2}$ in eq.(17).  This
can also be seen in Fig.1, which shows contours of $\log\bar\sigma$.

The fitting formula is quite successful.  For half of our measurements
the result is accurate to better than 10\%, and for about 75\% of our
measurements it is better than 20\%.  Of the remaining measurements
there are some in which the disagreement exceeds 2 standard
deviations, and they are shown in Fig.2.  At each point the label
gives the relative error in the sense $(\bar\sigma_{th} -
\bar\sigma_{num})/\bar\sigma_{th}$, where $\bar\sigma_{th}$ and
$\bar\sigma_{num}$ are the theoretical value (i.e. from eq.(17)) and
the numerical value (from Table 1), respectively.  Thus a value of
$-1$ would mean that $\bar\sigma_{th} = 0.5\bar\sigma_{num}$, while
the extreme positive value of $0.69$ means that $\bar\sigma_{th}\simeq
3\bar\sigma_{num}$.  Note that the cross sections around this point
are very low (Fig.1), which has two consequences: first, that the
standard deviation of the numerical cross section is comparable with
the cross section itself, and, second, that exchange may be
unimportant in this regime.  Indeed for all but two of the points
plotted in Fig.2 the discrepancy between the formula and the numerical
data is less than three standard deviations.  The two exceptions are
the points labeled $0.53$ and $-0.75$.  It is also reassuring that
almost all the discrepant points are surrounded by data points (Fig.1)
where the agreement between the numerical and semi-theoretical results
is satisfactory by the above criteria.  Thus the errors are localised.
Nevertheless, it is evident that there are some significant and
systematic trends in this data, and the possible effect of these
discrepancies should be assessed in any application of the fitting
formula.

\bigskip\noindent
{\bf 5. Discussion and Conclusions}
\medskip\noindent
5.1 Comparison with Other Authors
\smallskip

Before summarising our findings, our immediate task will be a
quantitative comparison of our fitting formula with some existing
results in the literature.  The first of these that we shall examine
is the paper of Sigurdsson \& Phinney (1993).  They give results for
different speeds $V$, and we have chosen the data for the lowest
speed, since the validity of our conclusions is restricted to the case
of hard binaries.  Data in their Tables 3A and 3B have been
normalised, where necessary, to the cross section $\sigma$ (eq.(3)),
and collected in our Table 4.  Also included are results of our
fitting formula, eq.(17), again converted to $\sigma$.

Unfortunately Sigurdsson \& Phinney do not give estimates of the
errors of their results, and in most of their runs the initial
eccentricity of the binary was chosen to be zero, and so only an
informal comparison is possible.  The agreement is often quite good,
and almost always within a factor of two.  In the cases where the
disagreement is most serious it is probably attributable to the
different choice of initial eccentricity distribution.  For example,
in the equal-mass case our fitting formula agrees with our numerical
data to within 10\%.  In the case $m_2\ll m_1=m_3$, however, where
their result for ejection of the low-mass component falls below that
of our fitting formula,  Sigurdsson \& Phinney 
have missed a significant fraction of exchange encounters by too small
a choice of the maximum impact parameter (E.S. Phinney, pers. comm.).
The fact that each of their data points is a weighted average over a
{\it range} of speeds $V$ may complicate the comparison further.

Now we turn to data presented by Hills (1992) on the case in which $m_1$
= $m_2$.  The initial eccentricity in his experiments was again 0, but
we find that our results agree with Hills' to within about a factor of
two over the entire range for which he found exchange events,
i.e. $0.3\ltorder m_3/m_1\ltorder 10^4$.  Hills' result exceeds ours
except at the lowest mass ratio in this range, and the discussion at the
end of \S3.2 suggests that the different choices of initial eccentricity
provide a plausible explanation for this last fact.

Finally in this section we compare our results with those of Rappaport
\etal\ (1989).  They computed the exchange cross section, by numerical
scattering experiments, for a sequence of binary pulsars.  The
sequence is characterised either by the orbital period or the mass of
the low-mass companion of the neutron star.  The cross section was
computed for an environment containing a stellar population drawn from
dynamical models of two globular clusters.  Here we restrict attention
to their binary of shortest period (3 days), since our results are
restricted to the hard-binary regime, and to the model of $\omega$
Cen, for illustration.  The results of Rappaport \etal\ give the
dimensionless scattering cross sections $\Sigma/(\pi a^2) = 1.3$ and
$55$, for ejection of the neutron star and low-mass companion,
respectively. Typical uncertainties are about 40\% and 15\%,
respectively. 

For our comparison we have used eq.(17) for each of the ten components
in the stellar population listed by Rappaport \etal, and have summed
the contributions, account being taken of their relative number
density and velocity dispersions.  Expressed in terms of the quantity
computed by them, our results are $\Sigma/(\pi a^2) = 1.37$ and
$76.0$.  The agreement is acceptable, considering the typical errors
in all results, and the fact that the result of Rappaport \etal\
applies to an initially circular binary.  It also illustrates the
utility of our results, as the cross section could be obtained for any
reasonable stellar population with little extra work.

\medskip
\noindent
5.2 Conclusions
\smallskip

This paper is a contribution to the theory of three-body classical
gravitational scattering.  This is a large topic with an extensive
literature, but it is the application to the dynamics of globular star
clusters which has provided the focus for our work. From the point of
view of this application, one of the most important processes is
exchange, whereby an incoming star ejects one component of a binary
and forms a new binary with the other component.  In the context of
star clusters we are also mainly concerned with encounters with hard
binaries, which are too energetic for an encounter to break up the
system into three single stars.  Finally, this application dictates
the importance of understanding scattering in a system where the stars
may have quite widely differing masses.

The main result of this paper is a semi-analytical cross section for
exchange, in the hard-binary limit, for all possible masses.  It has
been derived partly from theoretical considerations and partly from
extensive new numerical data on scattering events.  The theory allowed
us to estimate the dependence of the cross section on the masses, in
various limiting cases, and the numerical data showed how the
theoretical results can be parameterised so as to provide a better
fit, including the cases where the masses are comparable.

The result is given in eq.(17) with coefficients in Table 3, and we
now summarise the way in which this information may be used.  Suppose
a star of mass $m_3$ approaches a binary with components of mass
$m_1$, $m_2$, and that its speed, while still at a large distance from
the binary, is $V$ relative to the binary.  Let the initial semi-major
axis of the binary be $a$.  Then the cross section for events in which
the component of mass $m_1$ is ejected, leaving a binary consisting of
the other two stars, can be computed in the following way.  First
compute $M_{12} = m_1 + m_2$, and also $M_{23}$, $M_{13}$ and
$M_{123}$, defined similarly.  Next, compute $\bar\sigma$ from
eq.(17), where $N = 3$, $\mu_1$ and $\mu_2$ are defined in eqs.(18)
and the coefficients are taken from Table 3.  (Note that the
exponential in eq.(17) is simply that of the cubic $a_{00} +
a_{10}\mu_1 + a_{01}\mu_2 + \ldots + a_{12}\mu_1\mu_2^2 +
a_{03}\mu_2^3$.) Then the required cross section is given by solving
eq.(4) for $\Sigma$.  This result is approximately valid provided that
the binary is hard, i.e. $v^2\ll1$, where $v$ is defined by eq.(1).

In astrophysical units this can all be summarised in the formulae
$$\eqalign{
\Sigma = 1.39\left({a\over0.1\au}\right)\left({10\kms\over
V}\right)^2\left({M_{123}\over
M_\odot}\right)\left({M_{23}\over M_{123}}\right)^{1/6}\left({m_3\over M_{13}}\right)^{7/2}\left({M_{123}\over
M_{12}}\right)^{1/3}\left({M_{13}\over M_{123}}\right)\times\cr
\times\  {\rm{\lr e}}^{3.70 + 7.49\mu_1 - 1.89\mu_2 -15.49\mu_1^2 -2.93\mu_1\mu_2 -
2.92\mu_2^2 
+ 3.07\mu_1^3 + 13.15\mu_1^2\mu_2 -5.23\mu_1\mu_2^2 +
3.12\mu_2^3}\ \au^2, \cr}\eqno(19)
$$
where $\mu_1 = m_1/M_{12}$ and $\mu_2 = m_3/M_{123}$.  The condition
that the binary is hard is
$$
0.011\left({V\over 10\kms}\right)^2\left({a\over 0.1{\rm
AU}}\right)\left({M_{12}m_3M_\odot\over m_1m_2M_{123}}\right)\ll1.\eqno(20)
$$
(The last factor is dimensionless, and the factor $M_\odot$ may be
omitted if the masses are expressed in units of the solar mass.)

For example, consider a binary pulsar consisting of a neutron star of
mass $1.4M_\odot$ and a white dwarf companion of mass $0.2M_\odot$ in
an orbit of period 10 days.  Suppose it encounters a $10M_\odot$ black
hole with a relative speed of $10\kms$.  Then $a\simeq 0.11$AU, $\mu_2
\simeq 0.862$, and the left side of eq.(20) evaluates to $0.06$.  To
compute the cross section for ejection of the neutron star we have
$\mu_1 = 0.875$, and so eq.(19) gives $\Sigma \simeq 120{\rm AU}^2$.
Surprisingly, perhaps, the cross section is not much smaller than that
for ejection of the low-mass companion; this is obtained by setting
$\mu_1 = 0.125$, and  is approximately
$200{\rm AU}^2$. Thus, of all encounters leading to exchange, roughly
one third lead to ejection of the companion of higher mass, for these
parameters.

Some cautionary remarks are now in order.  First, the cross section in
eq.(19) is a statistical result, and one of the averages that is
implicitly performed in our work is over the initial eccentricity of
the binary.  It is assumed to have the ``thermal'' distribution $f(e)
= 2e$.  The limited evidence available (\S5.1) suggests that this
affects the result by at most a factor of two, except in regimes where
exchange is probable only when the eccentricity is high.  From the
discussion of \S3.2 (which also applies with some changes to \S3.3),
these events are those in which $m_3\ll m_1$; in such circumstances
exchange is actually very rare anyway, and so this issue is unlikely
to be important.

The second precaution concerns the error in the fitting formula.  We
have found that it usually agrees with numerical data to better than
20\%, but that there are a few places where the error can apparently
exceed 50\%.  These are illustrated in Fig.2, and it may be advisable
to check whether events in these areas of parameter space are of
importance in a given application.

Finally, no-one who makes use of these results in applications will
need to be reminded that they apply to the point-mass approximation.
In some cases the results would be drastically different for stars of
finite radius.

\bigskip\noindent We thank the referee for his detailed and helpful
comments. This work was supported in part by NASA grant
NAGW-2559 and NSF grant AST-9308005.  DCH thanks the Institute for
Advanced Study, Princeton, for its hospitality while much of the
theoretical work in this paper was being carried out.

\par\vfill\eject

\bigskip\noindent
{\bf References}
\parindent=0pt
\medskip
Becker L., 1920, MNRAS, 80, 590

Fabian A. C., Pringle J. E., Rees, M. J., 1975, 172, 15P

Goodman J.,  Hut P., 1993, ApJ,  403, 271 (Paper V)

Heggie D.C., 1975, MNRAS, 173, 729

Heggie D.C., Hut P., 1993, ApJS, 85, 347 (Paper IV)

Hills J.G., 1975, AJ, 80, 809

Hills J.G., 1991, AJ, 102, 704

Hills J.G., 1992, AJ, 103, 1955

Hills J.G., Dissly R.W., 1989, AJ, 98, 1069

Hills J.G.,  Fullerton, L.W., 1980, AJ, 85, 1281

Hut P., 1983, ApJ, 268, 342 (Paper II)

Hut P., 1984, ApJS, 55, 301

Hut P., 1985, in Goodman J., Hut P., eds, Dynamics of Star Clusters, IAU
Symp 113. Reidel, Dordrecht, p.231

Hut P., 1993, ApJ, 403, 256 (Paper III)

Hut P., Bahcall J.N., 1983, ApJ, 268, 319 (Paper I)

Hut P., Makino, M. \& McMillan, S.L.W. 1995, ApJ, 443, L93

McMillan, S.L.W., \& Hut P. 1996, in preparation (Paper VI).

Marchal C., 1990, The Three-Body Problem. Elsevier, Amsterdam.

Rappaport S., Putney A., Verbunt F., 1989, ApJ, 345, 210

Rasio F.A., McMillan S.L.W.,  Hut P., 1995, ApJL, 438, 33

Sigurdsson S., Phinney E.S., 1993, ApJ, 415, 631

\par\vfill\eject

\centerline{Table 1}
\smallskip
\centerline{Numerical Exchange Cross Sections $\bar\sigma$}
\settabs 11\columns
\bigskip
\hrule
\smallskip
\+ $m_1$&$m_2$&$m_3$&Direct Exchange&&&&Resonant Exchange\cr
\+Star:&&&1&&2&&1&&2\cr
\smallskip
\+ 0.500& 0.500& 0.005& $\ast$&& $\ast$&& $\ast$&& $\ast$\cr
\+ 0.500& 0.500& 0.017& $\ast$&& $\ast$&& $\ast$&& $\ast$\cr
\+ 0.500& 0.500& 0.050& $\ast$&& $\ast$&& 0.010$\pm$ 0.010&& $\ast$\cr
\+ 0.500& 0.500& 0.167& 0.024$\pm$ 0.012&& 0.024$\pm$ 0.012&& 0.072$\pm$ 0.024&& 0.114$\pm$ 0.042\cr
\+ 0.500& 0.500& 0.250& 0.015$\pm$ 0.008&& 0.036$\pm$ 0.016&& 0.326$\pm$ 0.058&& 0.310$\pm$ 0.060\cr
\+ 0.500& 0.500& 0.333& 0.068$\pm$ 0.017&& 0.093$\pm$ 0.021&& 0.975$\pm$ 0.135&& 1.065$\pm$ 0.134\cr
\+ 0.500& 0.500& 0.500& 0.357$\pm$ 0.032&& 0.330$\pm$ 0.028&& 1.962$\pm$ 0.119&& 1.883$\pm$ 0.121\cr
\+ 0.500& 0.500& 1.000& 1.202$\pm$ 0.076&& 1.165$\pm$ 0.069&& 3.128$\pm$ 0.176&& 2.969$\pm$ 0.169\cr
\+ 0.500& 0.500& 1.500& 1.657$\pm$ 0.139&& 1.569$\pm$ 0.128&& 3.625$\pm$ 0.297&& 3.304$\pm$ 0.278\cr
\+ 0.500& 0.500& 3.000& 2.425$\pm$ 0.263&& 2.867$\pm$ 0.325&& 3.508$\pm$ 0.427&& 3.539$\pm$ 0.447\cr
\+ 0.500& 0.500& 5.000& 3.134$\pm$ 0.343&& 4.348$\pm$ 0.460&& 4.390$\pm$ 0.526&& 3.965$\pm$ 0.525\cr
\+ 0.500& 0.500& 6.000& 4.607$\pm$ 0.372&& 3.796$\pm$ 0.327&& 4.070$\pm$ 0.422&& 4.396$\pm$ 0.470\cr
\+ 0.500& 0.500&15.000& 6.619$\pm$ 0.467&& 5.919$\pm$ 0.433&& 5.001$\pm$ 0.500&& 4.531$\pm$ 0.433\cr
\+ 0.500& 0.500&50.000&10.065$\pm$ 0.669&& 9.925$\pm$ 0.668&& 7.204$\pm$ 0.737&& 7.505$\pm$ 0.785\cr
\+ 0.500& 0.500&99.000& 8.594$\pm$ 1.948&&12.504$\pm$ 2.805&&14.804$\pm$ 4.737&&11.374$\pm$ 4.133\cr
\smallskip
\+ 0.333& 0.667& 0.667& 1.095$\pm$ 0.092&& 0.339$\pm$ 0.045&& 4.464$\pm$ 0.300&& 0.582$\pm$ 0.109\cr
\smallskip
\+ 0.400& 0.600& 0.400& 0.383$\pm$ 0.076&& 0.140$\pm$ 0.037&& 1.972$\pm$ 0.221&& 0.583$\pm$ 0.121\cr
\smallskip
\+ 0.250& 0.750& 0.007& $\ast$&& $\ast$&& $\ast$&& $\ast$\cr
\+ 0.250& 0.750& 0.025& $\ast$&& $\ast$&& $\ast$&& $\ast$\cr
\+ 0.250& 0.750& 0.075& 0.010$\pm$ 0.010&& $\ast$&& 0.040$\pm$ 0.020&& $\ast$\cr
\+ 0.250& 0.750& 0.083& $\ast$&& $\ast$&& 0.099$\pm$ 0.014&& $\ast$\cr
\+ 0.250& 0.750& 0.250& 0.276$\pm$ 0.039&& 0.009$\pm$ 0.005&& 2.189$\pm$ 0.165&& 0.063$\pm$ 0.024\cr
\+ 0.250& 0.750& 0.500& 1.087$\pm$ 0.094&& 0.065$\pm$ 0.016&& 4.856$\pm$ 0.302&& 0.115$\pm$ 0.042\cr
\+ 0.250& 0.750& 0.750& 1.359$\pm$ 0.095&& 0.298$\pm$ 0.042&& 5.721$\pm$ 0.318&& 0.431$\pm$ 0.101\cr
\+ 0.250& 0.750& 2.250& 2.863$\pm$ 0.088&& 1.840$\pm$ 0.077&& 5.606$\pm$ 0.167&& 1.358$\pm$ 0.093\cr
\+ 0.250& 0.750& 7.500& 4.349$\pm$ 2.050&& 2.626$\pm$ 1.700&& 6.129$\pm$ 3.700&& 5.378$\pm$ 3.650\cr
\+ 0.250& 0.750&22.500& 7.976$\pm$ 0.517&& 7.251$\pm$ 0.533&& 7.150$\pm$ 0.600&& 4.501$\pm$ 0.583\cr
\+ 0.250& 0.750&99.000&13.916$\pm$ 3.172&& 9.681$\pm$ 2.261&&14.478$\pm$ 4.345&&11.647$\pm$ 4.554\cr
\smallskip
\+ 0.100& 0.900& 0.200& 0.315$\pm$ 0.050&& $\ast$&& 3.038$\pm$ 0.218&& $\ast$\cr
\+ 0.100& 0.900& 1.000& 2.058$\pm$ 0.380&& 0.606$\pm$ 0.358&& 4.697$\pm$ 0.823&& 0.476$\pm$ 0.364\cr
\+ 0.100& 0.900& 2.000& 2.644$\pm$ 0.243&& 1.004$\pm$ 0.151&& 6.224$\pm$ 0.538&& 1.258$\pm$ 0.292\cr
\smallskip
\+ 0.091& 0.909& 0.009& $\ast$&& $\ast$&& $\ast$&& $\ast$\cr
\+ 0.091& 0.909& 0.030& 0.011$\pm$ 0.011&& $\ast$&& 0.109$\pm$ 0.038&& $\ast$\cr
\+ 0.091& 0.909& 0.091& 0.051$\pm$ 0.015&& $\ast$&& 1.035$\pm$ 0.104&& $\ast$\cr
\+ 0.091& 0.909& 0.303& 0.658$\pm$ 0.045&& 0.009$\pm$ 0.004&& 4.116$\pm$ 0.165&& 0.007$\pm$ 0.004\cr
\+ 0.091& 0.909& 0.909& 2.031$\pm$ 0.131&& 0.294$\pm$ 0.051&& 5.710$\pm$ 0.335&& 0.219$\pm$ 0.070\cr
\+ 0.091& 0.909& 2.727& 3.286$\pm$ 0.214&& 1.417$\pm$ 0.136&& 5.905$\pm$ 0.387&& 1.216$\pm$ 0.216\cr
\+ 0.091& 0.909& 9.091& 5.257$\pm$ 0.345&& 4.693$\pm$ 0.382&& 6.671$\pm$ 0.509&& 3.454$\pm$ 0.436\cr
\smallskip
\+ 0.050& 0.950& 0.025& $\ast$&& $\ast$&& 0.099$\pm$ 0.049&& $\ast$\cr
\+ 0.050& 0.950& 0.050& $\ast$&& $\ast$&& 0.827$\pm$ 0.186&& $\ast$\cr
\smallskip
\+ 0.032& 0.968& 0.010& $\ast$&& $\ast$&& $\ast$&& $\ast$\cr
\+ 0.032& 0.968& 0.032& 0.008$\pm$ 0.006&& $\ast$&& 0.457$\pm$ 0.072&& $\ast$\cr
\+ 0.032& 0.968& 0.097& 0.093$\pm$ 0.028&& $\ast$&& 2.009$\pm$ 0.146&& $\ast$\cr
\+ 0.032& 0.968& 0.323& 1.000$\pm$ 0.087&& $\ast$&& 4.294$\pm$ 0.250&& 0.012$\pm$ 0.012\cr
\+ 0.032& 0.968& 0.968& 2.264$\pm$ 0.145&& 0.345$\pm$ 0.058&& 5.445$\pm$ 0.332&& 0.232$\pm$ 0.069\cr
\+ 0.032& 0.968& 2.903& 3.703$\pm$ 0.237&& 2.210$\pm$ 0.213&& 6.178$\pm$ 0.409&& 1.629$\pm$ 0.258\cr
\smallskip
\+ 0.010& 0.990& 0.010& $\ast$&& $\ast$&& 0.091$\pm$ 0.022&& $\ast$\cr
\+ 0.010& 0.990& 0.033& 0.008$\pm$ 0.005&& $\ast$&& 0.850$\pm$ 0.097&& $\ast$\cr
\+ 0.010& 0.990& 0.099& 0.071$\pm$ 0.020&& $\ast$&& 2.095$\pm$ 0.150&& $\ast$\cr
\+ 0.010& 0.990& 0.330& 0.981$\pm$ 0.085&& 0.003$\pm$ 0.002&& 4.002$\pm$ 0.232&& 0.002$\pm$ 0.002\cr
\+ 0.010& 0.990& 0.990& 2.349$\pm$ 0.143&& 0.271$\pm$ 0.056&& 6.024$\pm$ 0.337&& 0.244$\pm$ 0.087\cr
\medskip\noindent
Note: the columns headed ``1'' and ``2'' give the cross sections for
exchange in which the particle of mass $m_1$ or $m_2$, respectively, is
ejected.
\bigskip
\bigskip
\centerline{Table 2}
\medskip
\centerline{Example of Discrepant Results}
\medskip
\settabs 5\columns
\+	&		&	&~~~~~~~~~~~~~  $\log_{10}\bar\sigma$\cr
\smallskip
\+$m_1$	&$m_2$		&$m_3$	&Formula	&Table 1\cr
\smallskip
\+0.99	&0.0099		&0.33	&-0.65		&-2.30$\pm$0.25	\cr
\+0.97  &  0.032  	&0.32   & -0.65		&-1.92$\pm$0.43	\cr
\+0.91  &  0.091	&0.30	&-0.65		&-1.80$\pm$0.15  \cr
\bigskip
\bigskip
\centerline{Table 3}
\medskip
\centerline{Coefficients for a Semi-Numerical Exchange Cross Section}
\medskip
\settabs 6\columns
\+&~~	$m$	&0	&1	&2	&3 \cr
\+&$n$	\cr
\+&0		&\phantom{-}3.70	&\phantom{-}7.49	&-15.49	&3.07\cr
\+&1		&-1.89	&-2.93	&\phantom{-}13.15\cr
\+&2		&-2.92	&-5.23\cr
\+&3		&\phantom{-}3.12\cr

\par\vfill\eject

\centerline{Table 4}

\smallskip
\centerline{Comparison with Results of Sigurdsson \& Phinney for $\sigma$}
\medskip
\settabs 7\columns
\+$m_1$&$m_2$&$m_3$&Sigurdsson \& Phinney&&This paper\cr
\+&&&Star 1&Star 2&Star 1&Star 2\cr
\smallskip
\+     1.0&     0.8&     1.0&     1.5&     2.8&     1.88&     3.37\cr
\+     1.0&     0.4&     1.0&     1.2&     7.0&     1.67&    11.2\cr
\+     1.0&     0.2&     1.0&     1.6&    15&     1.85&    23.9\cr
\+     1.0&     0.1&     1.0&     2.6&    28&     2.61&    44.1\cr
\+     1.0&     0.05&     1.0&     5.0&    50&     4.35&    80.8\cr
\+     1.0&     0.025&     1.0&    10&   100&     7.93&   152\cr
\+     1.0&     0.0125&     1.0&    20&   170&    15.2&   295\cr
\+     1.0&     1.0&     0.4&     0.060&     0.060&     0.128&     0.128\cr
\+     1.0&     1.0&     1.0&     1.1&     1.1&     1.96&     1.96\cr
\+     0.5&     0.35&     1.0&     5.5&     9.8&     8.07&    14.4\cr
\+     1.0&     0.35&     0.5&     0.1&     2.8&     0.183&     3.78\cr
\medskip\noindent
Note: the columns headed ``Star 1'' and ``Star 2'' give the cross
sections for exchange in which the particle of mass $m_1$ or $m_2$,
respectively, is ejected.

\vfill\eject
\parindent=20pt
\noindent
{\bf Appendix. Detailed Balance for Exchange Cross Sections}
\medskip
The theory of detailed balance is described in some generality in
Heggie (1975), though it is expressed in terms of rate functions,
i.e. the integral of a cross section over a Maxwellian distribution of
velocities, and does not explicitly deal with exchange reactions.
Detailed balance is also described in terms of cross sections in Paper
III in this series, though in the case of equal masses.  Since the
integration over a Maxwellian is essentially a Laplace transform, it
is possible to obtain the result for cross sections from the result in
Heggie (1975), and this will be the starting point for the following
treatment.  We have verified that a direct derivation for cross
sections for exchange reactions with different masses leads to the
same result.

With some changes of notation the result presented in Heggie (1975)
can be written as 
$$\eqalign{
 {1\over2}n_1n_2n_3(\pi/kT)^{3/2}(m_1m_2)^3\vert
E_{12}\vert^{-5/2}{dR\over dE^\prime_{12}}(E_{12}\to
E^\prime_{12})\exp(- E_{12}/kT) =&\cr
{1\over2}n_1n_2n_3(\pi/kT)^{3/2}(m_1m_2)^3\vert
E^\prime_{12}\vert^{-5/2}{dR\over dE_{12}}(E^\prime_{12}\to
E_{12})&\exp(- E^\prime_{12}/kT),\cr}\eqno(A.1)
$$ 
where $dR(E_{12}\to E^\prime_{12})$ is the rate (per unit density of
reactants) of reactions which change the binding energy of a binary
from $E_{12}$ to a value $E^\prime_{12}$ within a range of size
$dE^\prime_{12}$, $T$ is the kinetic temperature, and $n_i$ is the
number density of stars with mass $m_i$; these cancel from this
equation, and are irrelevant in what follows.  Eq.(A.1) is appropriate
to encounters which do not lead to exchange, and the use of the labels
$12$, which identify the components of the binary, seems pedantic at
this stage, but it becomes useful when we go on to discuss exchange.

Now the rate function can be defined in terms of a differential cross
section by
$$
{dR\over dE^\prime_{12}}(E_{12}\to
E^\prime_{12}) =
 \left({2\pi kTM_{123}\over
m_3M_{12}}\right)^{-3/2}\int
V_3\exp\left(-{m_3M_{12}V_3^2\over2M_{123}kT}\right){d\Sigma\over dE^\prime_{12}}(E_{12}\to
E^\prime_{12})d^3\bV_3.
$$
Then we can find the detailed balance relation for the differential
cross section by substituting this integral into eq.(A.1) and
inserting a delta function
$\displaystyle{\delta\left({m_3M_{12}\over2M_{123}}V_3^2 + E_{12} -
E\right)}$, which isolates interactions involving systems with total
energy $E$ in their barycentric frame.  (Primed variables are used on
the right side of eq.(A.1).)  Cancelling symmetric functions of the
masses, the result we obtain is
$$\eqalign{
(m_1m_2)^3(m_3M_{12})^{1/2}\vert E_{12}\vert^{-5/2}V_3^2{d\Sigma\over dE^\prime_{12}}(E_{12}\to
E^\prime_{12}) &= \cr
(m_1m_2)^3(m_3M_{12})^{1/2}\vert E^\prime_{12}\vert^{-5/2}V^{\prime^2}_3&{d\Sigma\over dE_{12}}(E^\prime_{12}\to
E_{12}).\cr}
$$

Now we observe that a similar relation holds for exchange reactions,
provided that the masses are correctly identified.  Thus
$$\eqalign{
(m_1m_2)^3(m_3M_{12})^{1/2}\vert E_{12}\vert^{-5/2}V_3^2{d\Sigma\over dE^\prime_{23}}(E_{12}\to
E^\prime_{23}) =&\cr
(m_2m_3)^3(m_1M_{23})^{1/2}\vert E^\prime_{23}\vert^{-5/2}V^{\prime^2}_1&{d\Sigma\over dE_{12}}(E^\prime_{23}\to
E_{12}).\cr}
$$
We can now drop the primes as the start and end states are
sufficiently identified by the subscripts, and we deduce that
$$
{d\Sigma\over dE_{12}}(E_{23}\to
E_{12}) = \left({m_1\over m_3}\right)^{5/2}\left({M_{12}\over
M_{23}}\right)^{1/2}{V_3^2\over V_1^2}\left({E_{12}\over E_{23}}\right)^{-5/2}
{d\Sigma\over dE_{23}}(E_{12}\to
E_{23}),
$$
which is eq.(11) in this paper.

\par\vfill\eject

\noindent
{\bf Figure Captions}

\medskip\noindent
{\bf Figure 1:}~~Coverage of the parameter space of mass ratios in the
numerical experiments.  Open circles represent experiments where the
cross section was too small to be measurable.  In this figure, $m_1$
is the mass of the component which is ejected.  Dashed lines are
contours of the logarithm of the theoretical exchange cross section
$\log_{10}\bar\sigma$ given by eq.(17).  The values of
$\log_{10}\bar\sigma$ range from $-5$ at lower right in steps of $0.5$
to $1$.

\medskip\noindent
{\bf Figure 2:}~~Data points where the fit of the semi-analytical
formula, eq.(17), is relatively poor.  At each value of the mass ratio
where the relative error exceeds both 20\% and two standard
deviations, the relative error is printed.

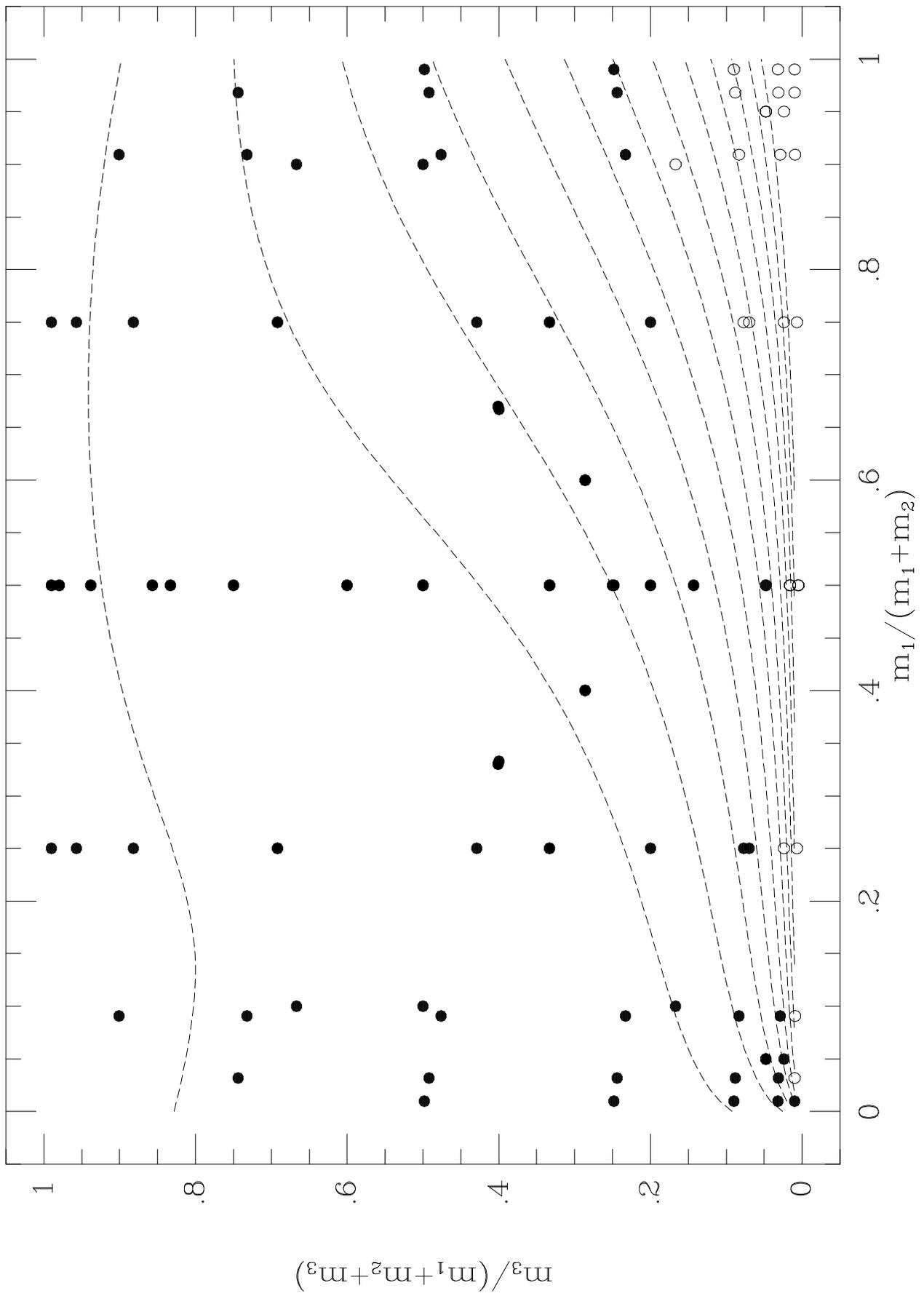
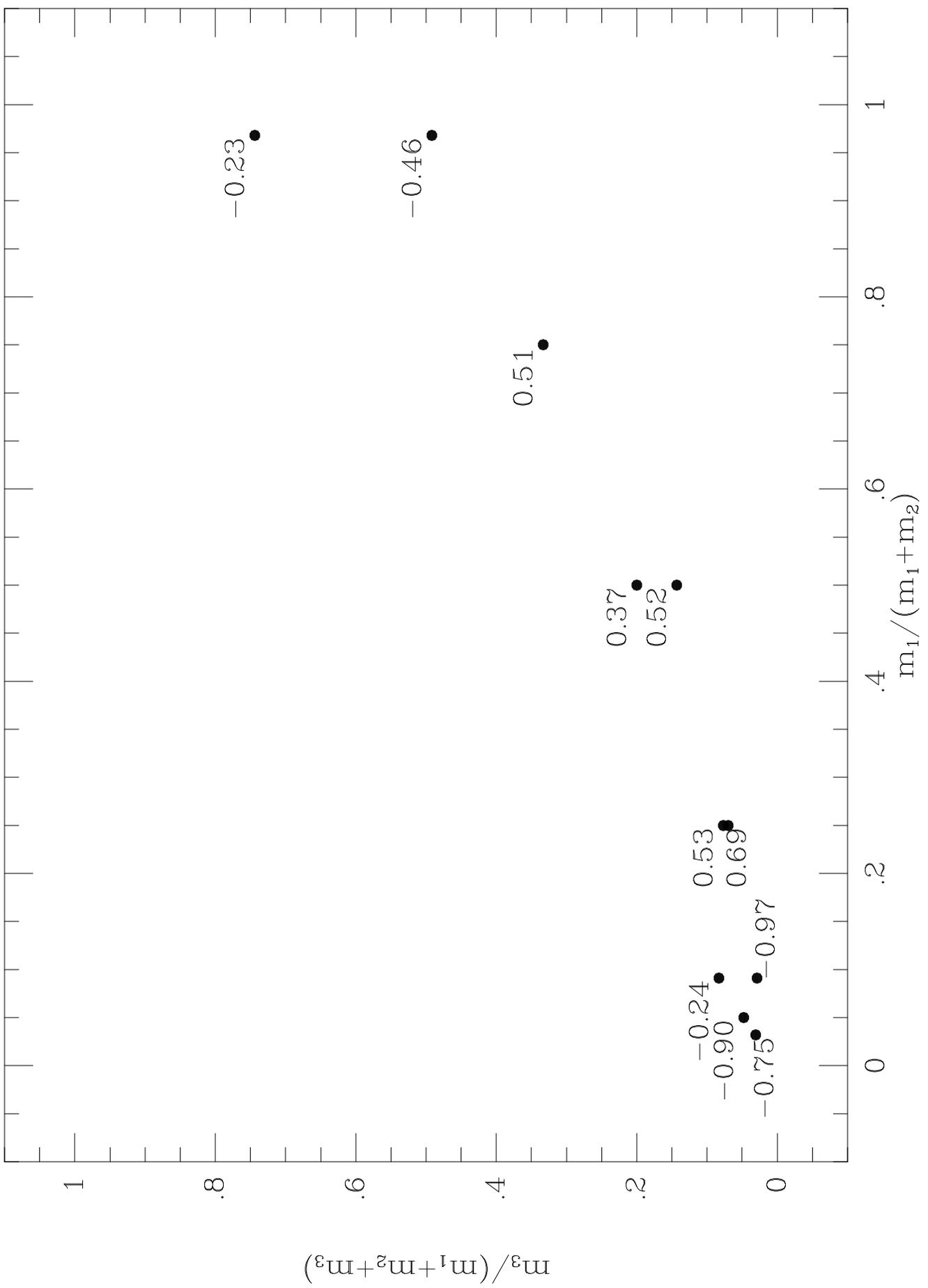
\bye